\documentclass[aps,prl,amsmath,amssymb,twocolumn,groupedaddress,showpacs]{revtex4-1}

\usepackage{graphicx}
\usepackage{dcolumn}
\usepackage{bm}

\usepackage{color}

\usepackage{textcomp}
\newcommand{\micm}{\,\text{\textmu m}}

\begin{document}

\preprint{APS/123-QED}

\title{Optical eigenmode imaging}

\author{Anna Chiara De Luca}
\author{Sebastian Kosmeier}
\author{Kishan Dholakia}
\author{Michael Mazilu}
\email{michael.mazilu@st-andrews.ac.uk} 
\homepage{\\ http://www.st-andrews.ac.uk/$\sim$atomtrap/}
\affiliation{SUPA-School of Physics and Astronomy, University of St Andrews, North Haugh, KY16 9SS, St Andrews, UK}

\date{\today}

\begin{abstract}
We present an indirect imaging method that measures both amplitude and phase information from a transmissive target. Our method is based on an optical eigenmode decomposition of the light intensity and the first-order cross correlation between a target field and these eigenmodes.  We demonstrate that such optical eigenmode imaging does not need any \emph{a priori} knowledge of the imaging system and corresponds to a compressive full-field sampling leading to high image extraction efficiencies. Finally, we discuss the implications with respect to second-order correlation imaging. 
\end{abstract}


\maketitle

The last two decades have seen the emergence of interest in schemes for 
interaction-free imaging~\cite{White:1998p1303} and 
indirect imaging~\cite{Bennink:2002p1254} in both the quantum and classical domain~\cite{Erkmen:2010p1246}. Such schemes inherently use light fields to image that  have themselves never scattered from the object. Prominent amongst these schemes has been the concept of ghost imaging~\cite{Ferri:2005p1255, Scarcelli:2006p1259, Shapiro:2008p1305, Bromberg:2009p1258}. These methods allow the image of an unknown object to be nonlocally reconstructed by the intensity correlation measurements between two light fields. The light that illuminates the object is typically collected by a single pixel detector that itself has no spatial resolution. The underlying physics of ghost imaging has seen major debate about its classical and quantum implications as both entangled source and thermal light can be used. Whilst a powerful concept, ghost imaging does not inherently reveal phase information about an object as it relies on a second-order correlation for its implementation~\cite{Bennink:2002p1254,Gong:2010p10137}.  Additionally, the resolution of the recovered images is determined by the size of the speckle placed on the object plane~\cite{Ferri:2005p1255}. Any imperfections or aberrations within the system may not be readily dealt with in such a system. A significant step forward would be an indirect imaging scheme based on correlation measurements that could retrieve both amplitude and phase and inherently accommodate issues relating to aberrations or imperfections within the optical system 

In this paper, we realise such a scheme using the concept of optical eigenmodes ~\cite{Mazilu:2009p1247,Mazilu:2011p1220}.
We split the laser light in two different beams. One beam does not interact with the target, but illuminates a high-resolution CCD camera (multi-pixel detector). The other one interrogates, in transmission, the target (or sample) and then illuminates a photodiode (single-pixel detector) providing no spatial resolution. The transmission wave-front of this beam is decomposed, using an optical lock-in amplification technique, onto an orthogonal set of optical eigenmodes. The lock-in amplification corresponds to performing a first order cross-correlation and as such is distinct from present ghost-imaging techniques. In turn this leads to the retention of the phase information of the object. Further, our approach foregoes the point-by-point scanning and allows a rapid full field image extraction~\cite{Herman:2009p1260}.

The outline of the paper is as follows. Firstly, the concept of the optical eigenmodes is briefly reviewed and then we discuss the problem of combining eigenmodes and indirect imaging. We introduce the experiment and show an application of the eigenmode method for indirect imaging of a target. We demonstrate the advantage of optical eigenmode imaging in terms of resolution and phase information. Finally, we show the relationship between optical eigenmodes and the second-order correlation function.

{\it Method.} To define the intensity optical eigenmodes~\cite{Mazilu:2011p1220}, we decompose a linearly polarised electromagnetic field $\textrm{E}$ into a superposition of $N$ monochromatic ($e^{i\omega t}$) ``test'' fields:
\begin{eqnarray}
{\rm E}=\sum_ja^*_j {\rm E}_j\;\;\; ; \;\;\;
{\rm E}^*=\sum_k{\rm E}^*_k a_k.
\label{E}
\end{eqnarray}
Here, we consider the field intensity $\it{m}^{(I)}$ integrated over a region of interest (ROI), as defined by:
\begin{equation}
m^{(I)}( {E})=\int _\text{ROI} d\sigma \;{\rm E}\cdot {\rm E}^*
\label{Intensity}
\end{equation}
valid for linearly polarised light and not tightly-focused beams. The ROI represents the detector active area. Equation (\ref{Intensity}) can be written in a general quadratic matrix form:
\begin{equation}
m^{(I)}({\rm E})=\sum_{j,k} a^*_j M_{jk}a_k
\label{Intensity2}
\end{equation}
where the elements $M_{jk}$ are constructed by combining the fields ${\rm E}_j$ and ${\rm E}_k$ for $ j,k=1...N$:
\begin{eqnarray}
M_{jk}=\int_\text{ROI} d\sigma {\rm E}_j\cdot{\rm E}^*_k .
\label{matrix}
\end{eqnarray}
The optical eigenmodes are defined by:
\begin{eqnarray}
 {\mathbb{E}_\ell} =\frac{1}{\sqrt{\lambda^\ell}} \sum_j v^*_{\ell j}{\rm E}_j \;\;\; ; \;\;\;
{\mathbb{E}^*_\ell} =\frac{1}{\sqrt{\lambda^\ell}}\sum_j {v}_{\ell j}{\rm E}_j^*
 \label{OE2}
\end{eqnarray}
with 
\begin{eqnarray}
 \sum_j M_{jk}{ v}_{\ell j}=\lambda^\ell { v}_{\ell k}
 \label{OE}
\end{eqnarray}
where $\lambda^\ell$ is an eigenvalue  and ${v}_{\ell j}$ the associated eigenvector.
The matrix  $M_{jk}$ shows two important properties.
Firstly, it is Hermitian meaning that all the eigenvalues ($\lambda^\ell$) are real and can be ordered where $\lambda^{\ell=1}$ is the largest eigenvalue,  $\lambda^{\ell=2}$ the second largest eigenvalue, etc... .
Secondly, two eigenvectors corresponding to different eigenvalues are orthogonal which means that: $ \int_\text{ROI}d\sigma\; {{\mathbb{E}}_j\cdot\mathbb{E}}^*_k=\delta_{jk}$.
As matter of fact, each of these eigenvectors ${v}_{\ell j}$  corresponds to a superposition of the initially considered fields. 
An unknown field ${\rm T}$ can be decomposed onto the eigenmodes using its projection defined by:
\begin{eqnarray}
c_\ell^*=\int_\text{ROI}d\sigma \;{ \rm T}\cdot {\mathbb{E}}^*_\ell .
 \label{projection}
\end{eqnarray}
where $c_\ell$ corresponds to the complex decomposition coefficients of the field ${ \rm T}$ in base $\mathbb{E}_\ell$. 
If the $\mathbb{E}_\ell$ fields form a complete base, we can perfectly reconstruct the unknown field ${\rm T}$ from the projection using ${\rm T}=c^*_\ell  {\mathbb{E}_\ell}$. We remark that the completeness of the base is dependent on the initial fields probing all the degrees of freedom available.  

Optical eigenmode imaging is based upon a direct measure of the projection coefficients $c^*_\ell$ and the experimental reconstruction of the unknown field  ${ \rm T}$, that is essentially the transmission function through the target.\\
\begin{figure}
\includegraphics[width=7.5cm]{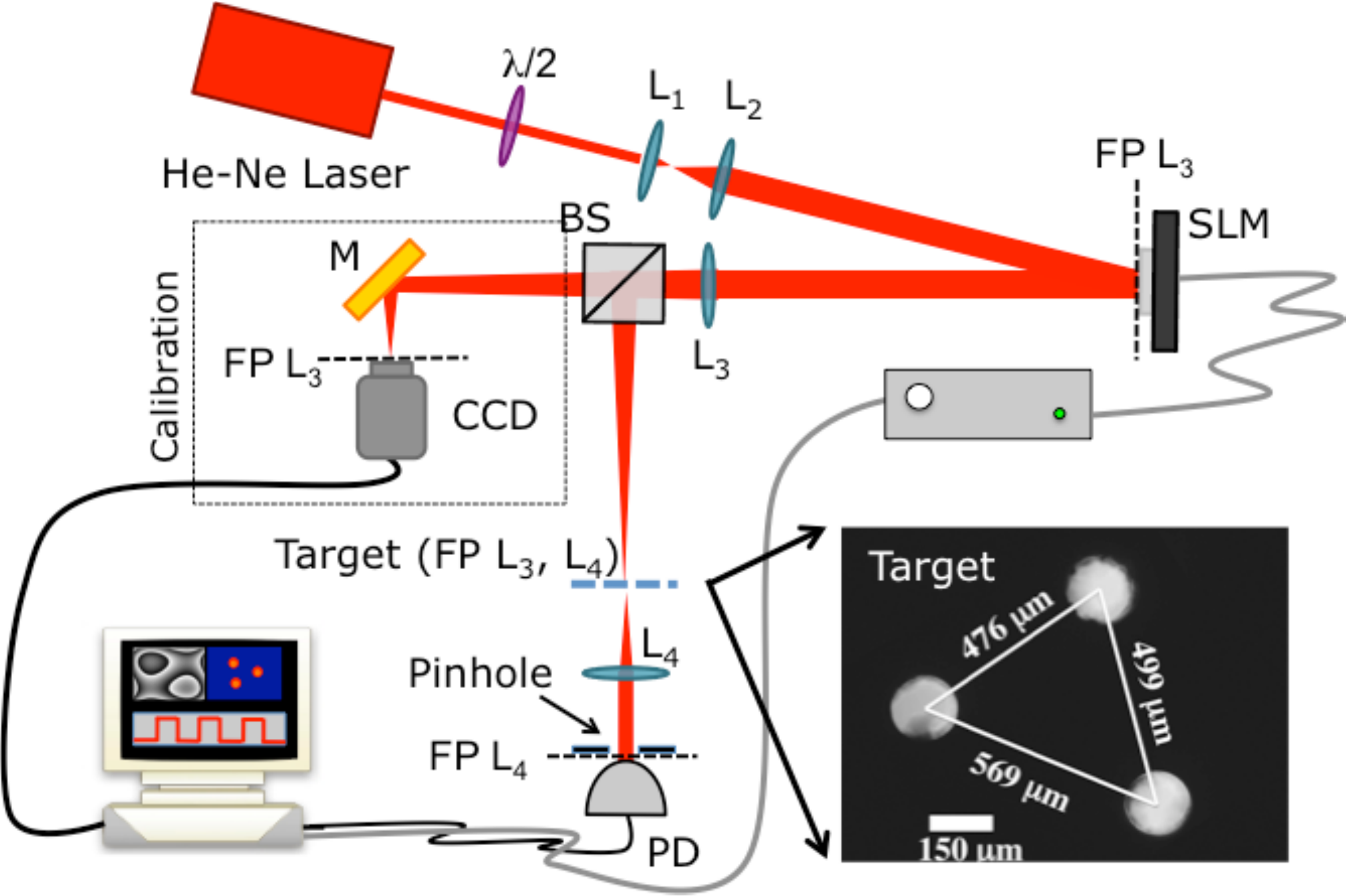}
\caption{\label{fig1} The experimental set-up used to perform  indirect imaging with the optical eigenmode formalism. The inset shows a scheme of the target of three holes used in our experiment. Abbreviations: L=lens, M=mirror, BS=beam splitter, FP= focal plane, PD= photodiode.}
\end{figure}
{\it Experimental.} The basic set-up used for the optical eigenmode imaging is shown in Fig.~\ref{fig1}.
A He-Ne laser (Thorlabs, $\lambda$ = 633 nm, P$_{max}$ = 10 mW) beam is expanded to fill the aperture of the spatial light modulator, SLM (Hamamatsu LCOS-SLM X10468, resolution = 600 x 800 pixels, pixel size = 20 x 20 $\mu$m$^2$, Refresh rate = 60 Hz) and then phase-modulated in the first-order configuration~\cite{DiLeonardo:2007p1265}. This configuration allow us to achieve phase and intensity modulation using the interfering efficiency.
The phase-modulated beam is  divided in two symmetric parts by a 50:50 beam splitter. The reflected beam interrogates a transmissive target placed in the focal plane of the Fourier lens (L3=75cm). A second Fourier lens (L4) integrates the transmitted light on a photodiode providing no spatial resolution. The transmitted beam is focused directly on a high-resolution CCD camera (Basler pilot piA640-210mg, resolution = 648 x 488 pixels, pixel size = 7.4 x 7.4 $\mu$m$^2$) without interacting with the target. By projecting the optical eigenmodes on the target it is possible to reproduce the target image on the CCD camera.

The experiment consists of two different steps.
Firstly, it is necessary to determine the optical eigenmodes (${\mathbb{E}_\ell}$) in the CCD camera plane by interfering the test fields ${\rm E}_j$ created in the SLM plane. Next, we select the $n$ most intense optical eigenmodes and interfere them with a reference signal (${\rm E}_\text{ref}$). The reference signal (${\rm E}_\text{ref}$) is random phase encoded \cite{Spalding2008fk} on the same SLM and thus shares the same optical path as the test field. In this way, we can reconstruct the amplitude and phase of the linearly polarised fields ${\rm F}_{j}$ on the CCD camera: 
\begin{eqnarray}
{\rm F}_{j}=\frac{1}{R} \sum^{R-1}_{p=0} e^{i2 \pi p/R}|{\rm E}_\text{ref}+ e^{-i2\pi p/R}{\rm E}_{j}|^2
\label{matrix2}
\end{eqnarray}
where R is an integer number greater then 3.  This procedure can be seen as a phase sensitive lock-in technique where the reference beam ${\rm E}_\text{ref}$ corresponds to a reference signal with respect to which the phase and amplitude of ${\rm E}_j$ is measured. This is done by measuring R times the intensity on the CCD camera while changing the relative phase between the reference and the test field. This intensities are then combined in a equation (\ref{matrix2}) in an inverse Fourier transform delivering a measure proportional to the complex test field ${\rm E}_j$. This approach generalises the four points ($R=4$) method presented in~\cite{Mazilu:2011p1220}. We remark here that, due to the lock-in technique, using a larger number of measures R experimentally delivers a better signal to noise ratio. The intensity operator $M_{jk}$ is given by:
\begin{eqnarray}
M_{jk}=\int _\text{ROI} d\sigma{\rm F}_j \cdot {\rm F}^*_k
\label{matrix3}
\end{eqnarray}
where ROI is a region of interest on the CCD camera.

In the second experimental step, we project the optical eigenmode on the target (T) using a first-order cross-correlation with the reference beam. The projection coefficients $c_k$ are related to the signal $s_k$ measured by the single-pixel detector (PD) as follows: 
\begin{eqnarray}
s_{k}&=&\frac{1}{R} \sum^{R-1}_{p=0} e^{i2 \pi p/R}\left|\int _\text{ROI}d\sigma \; {\rm T} \cdot({\rm E}_\text{ref}+ e^{-i2\pi p/R}{\mathbb{E}_k})\right|^2 \nonumber \\
&=&  \int_\text{ROI} d\sigma \;  {\rm T} \cdot {\rm E}^*_\text{ref}   \cdot \int_\text{ROI}d\sigma \; {\rm T}^*  \cdot {\mathbb{E}_k}\nonumber \\
&=&c_k  \int_\text{ROI}d\sigma \;  {\rm T}\cdot {\rm E}^*_\text{ref} 
 \label{PD}
\end{eqnarray}
where $\int_\text{ROI} d\sigma \; {\rm T} \cdot {\rm E}^*_\text{ref} $ gives the coupling coefficient to the target and is essentially a  constant value. Therefore, from the signal acquired on the PD, we can measure the coefficients $c_k$ which makes it possible to reconstruct the complex target field. 
Finally, using a Dirac distribution target field, we can deduce the point spread function ${\rm E}_{\text{PSF}}$ which determines the resolution of the optical eigenmode imaging method:
\begin{eqnarray}
{\rm E}_{\text{PSF}}=\sum_j {\mathbb{E}}_j \int_\text{ROI}\delta(r-r_0) {\mathbb{E}}^*_j d\sigma
 \label{projection3}
\end{eqnarray}
\begin{figure}
\includegraphics[width=7 cm]{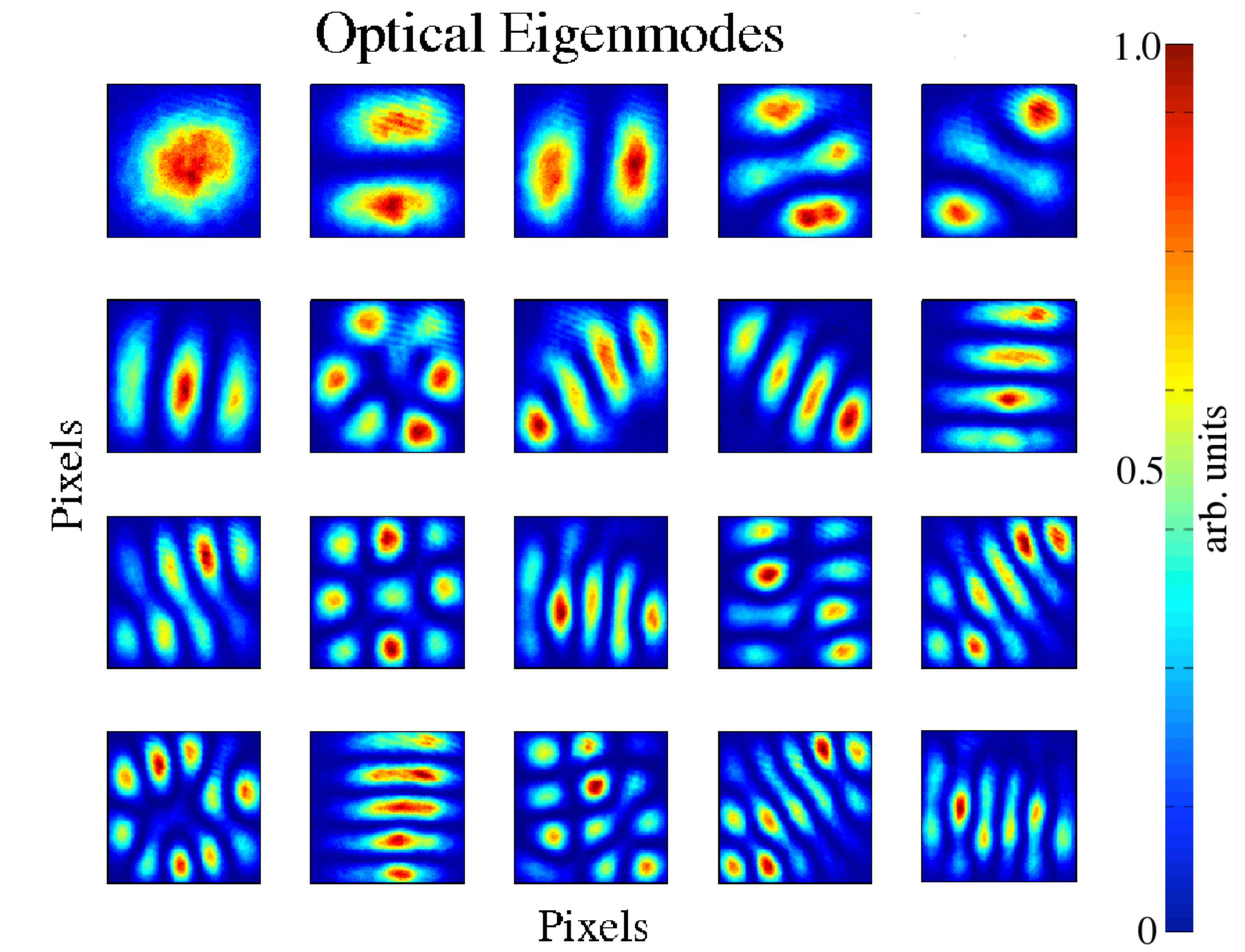}
\caption{\label{fig2} Examples of the principal experimental intensity optical eigenmodes. The resolution of each image is 120 x 120 pixels and the scale is not shown for simplicity.}
\end{figure}
{\it Results and discussion.} In our experiment, we used $N=1353$ masks consisting of 33 horizontal and 41 vertical independent deflection angles to determine the intensity optical eigenmodes (${\mathbb{E}_\ell}$) in the CCD camera plane.  After the Fourier lens (L3), these deflections correspond to a regular grid scan over an area of 1.2mm by 1.3mm. The reference field (${\rm E_\text{ref}}$)   was chosen to be the median beam out of the 1353 beams used. The first 20 experimental intensity optical eigenmodes are shown in Fig.~\ref{fig2} displaying visually an orthogonal behaviour similar to TEM modes. 

\begin{figure}
\includegraphics[width=7cm]{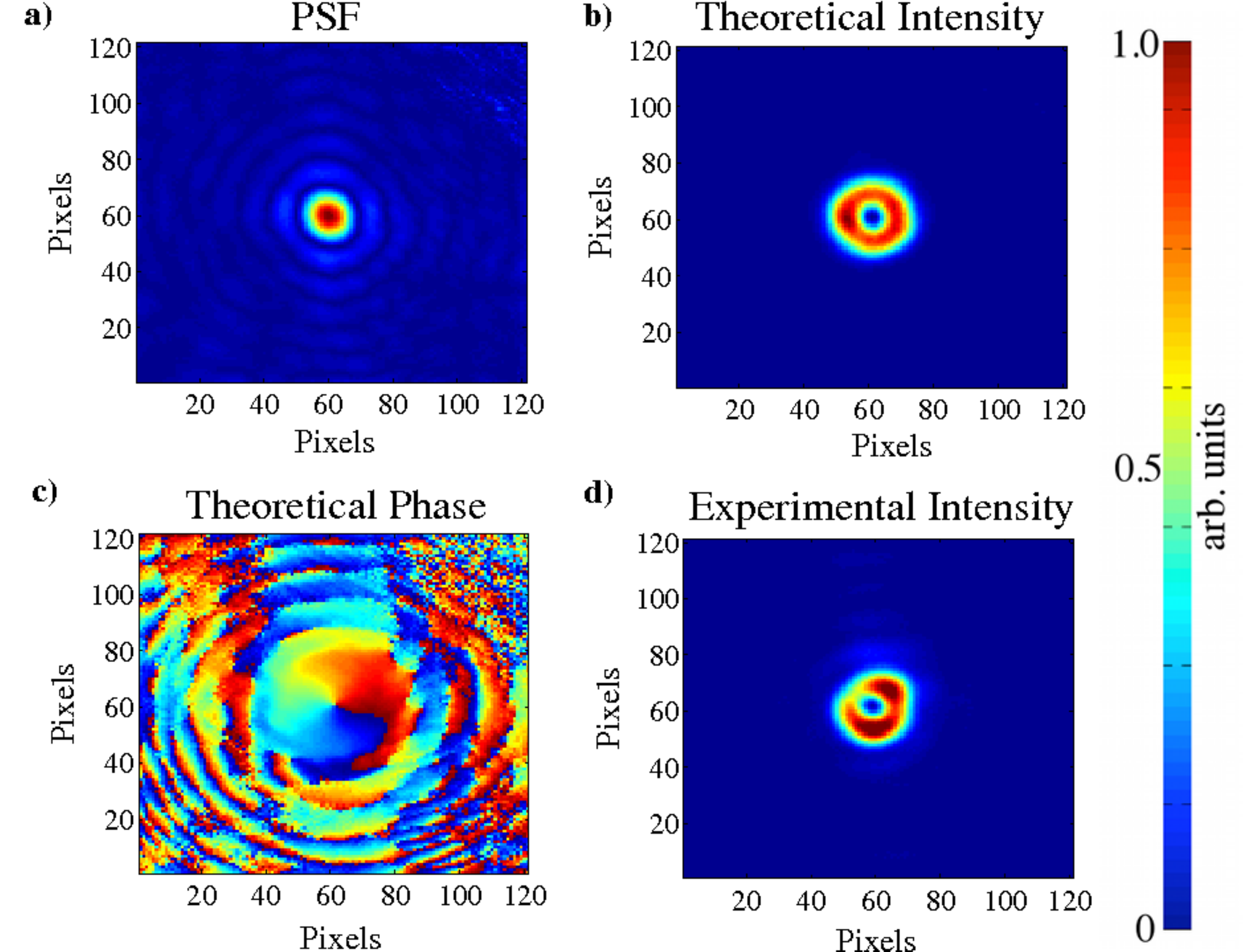}
\caption{\label{fig3} Experimental optical eigenmode Imaging of a Laguerre-Gauss (LG) beam. (a) Point Spread Function (PSF) obtained by combining 147 Intensity OEi. (b) Theoretical intensity  and (c) phase of an decomposed LG beam (Intensity scale bar is from 0 (blue) to 2$\pi$ (red)). (d) Experimental reconstruction of the LG beam using the 147 optical eigenmodes. }
\end{figure}

From the experimental intensity optical eigenmodes, we select the ones with an efficiency above 0.1\% with respect to the maximum possible, giving us typically around 140 eigenmodes. This corresponds to a tenfold image data reduction when compared to the initial number of test fields ($N=1353$).  Further, we noticed no clear increase in the number of suitable eigenmodes when increasing $N$. The quality of this optical compression technique can be assessed experimentally by determining the PSF (\ref{projection3}) showing the loss of image information when reconstructing a Dirac distribution field. Fig.~\ref{fig3}-a shows the experimental PSF obtained by combining 147 intensity optical eigenmodes representing the resolution limit of the system. Further, we numerically project the experimental eigenmode fields onto a simulated Laguerre-Gaussian target with a unit vortex charge. The theoretical reconstructed field is displayed in Fig.~\ref{fig3}-b and Fig.~\ref{fig3}-c as field intensity and phase respectively. Fig.~\ref{fig3}-d shows the intensity on the CCD camera when the SLM displays the optical eigenmode superposition corresponding to ${\rm T}=\sum_{\ell} c^*_\ell  {\mathbb{E}_\ell}$. These figures show that optical eigenmode imaging provides phase and intensity information as predicted in equation (\ref{projection}).

\begin{figure}
\includegraphics[width=7cm]{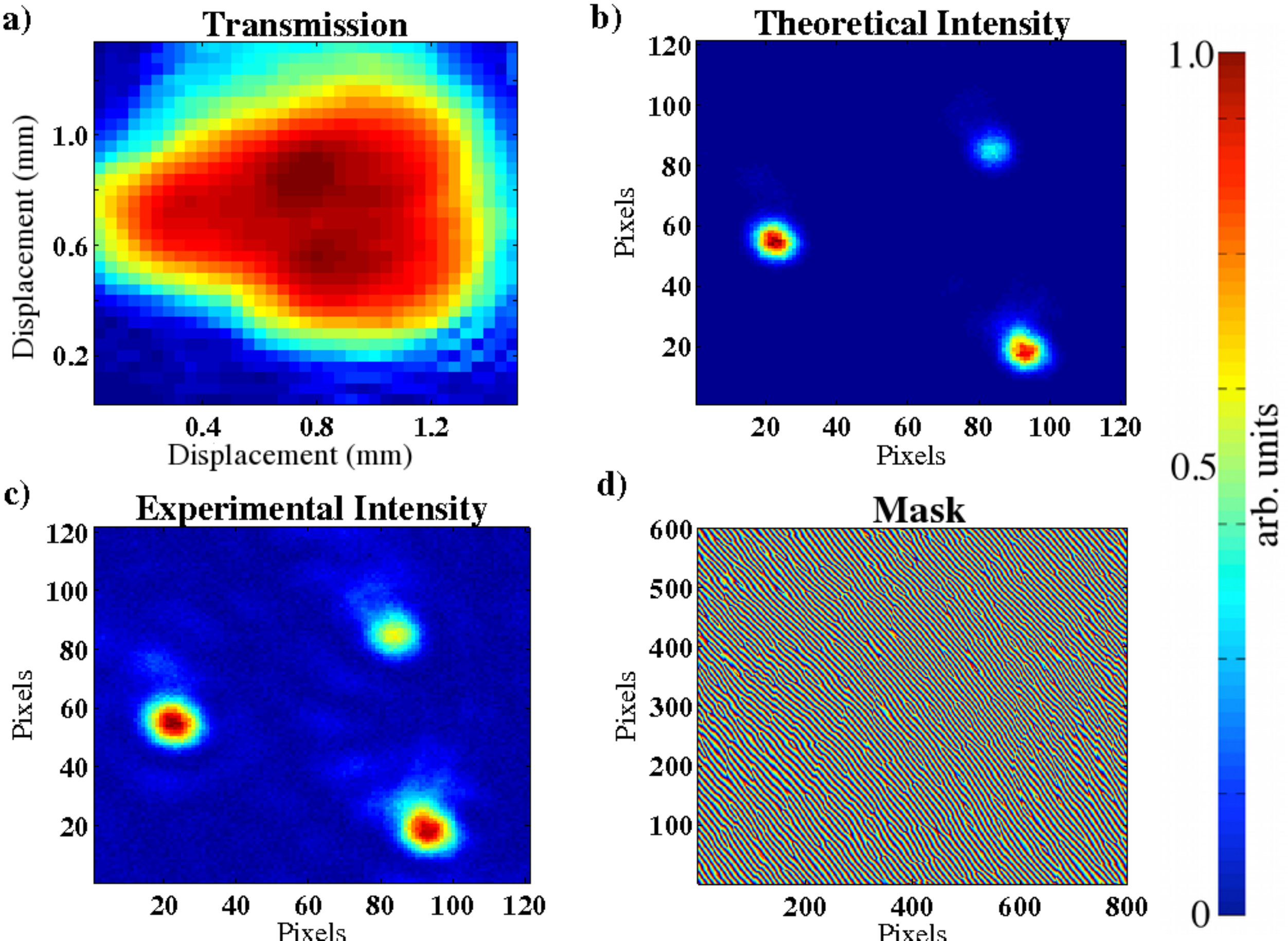}
\caption{\label{fig4} Indirect Imaging of a three holes target. (a) Conventional transmission image of the target reconstructed from the intensity signal collected by the PD as a function of the beam displacement in the target plane. (b) Corresponding numerical indirect intensity Image of the target; (c) Experimental indirect optical eigenmode Image; (d) Final mask encoded on the SLM.}
\end{figure}

For the indirect imaging we used a target consisting of three holes each with a diameter of $\sim$150$\micm$ (see Fig.\ref{fig1} for more details). Fig.~\ref{fig4}-a shows the  conventional transmission image of the target obtained by scanning the laser spot, using the initial 1353 deflection masks, through the holes.  In this figure, we plot the signal intensity detected by the single-pixel detector (PD) for each angle of the ligth scanning.  By contrast, Fig.~\ref{fig4}-b and Fig.~\ref{fig4}-c show the indirect image obtained, respectively through numerical and experimental optical eigenmode superpositions, on the CCD camera.  The latter was produced using the SLM mask reported in Fig.~\ref{fig4}-d.
Direct comparison of the theoretical superposition (Fig.~\ref{fig4}-b) and the experimental one (Fig.~\ref{fig4}-c) clearly reveals  a good agreement. These results suggest that the numerical and the experimental distributions verify: $s_k\propto c_k$, demonstrating the linearity of our optical system. 
Further, by comparing Fig.~\ref{fig4}-a and c, we can clearly observe an enhancement in the resolution of the optical eigenmode image with respect to the conventional transmission image using the initial scanning ``test'' fields.  

Remarkably, the optical eigenmodes described above simplify the first-order  correlation functions. Indeed, we have
\begin{eqnarray}
G^{(1)}(\tau)&=&\int _\text{ROI} \left<{\rm E}(t) {\rm  E}^{*}(t+\tau) \right>d\sigma\nonumber \\
&=&e^{-i \omega \tau} \left< \sum_{j,k} a_j^* M_{jk} a_k\right>=\sum_{jk}G^{(1)}_{jk}(\tau)
\label{g1}
\end{eqnarray}
with
\begin{eqnarray}
G^{(1)}_{jk}(\tau)=\int _\text{ROI} \left< \mathbb{ E}_j(t)  \mathbb{E}^{*}_k(t+\tau) \right>d\sigma=e^{-i \omega \tau}\delta_{jk}
\label{g1}
\end{eqnarray}
where the symbol $\rm{<...>}$ indicates the ensemble averaging. 
This relationship shows that the optical eigenmodes are independent with respect to the first-order correlation function i.e. amplitude and phase  fluctuations of two different  eigenmodes are not correlated. This implies that any random thermal fluctuations of the illumination source can be decomposed into independent fluctuations each corresponding to an eigenmode. We also remark that the phase information is maintained in the first order correlation process. It is these two properties that make the optical eigenmode indirect imaging possible. Additionally, illuminating the target directly with the eigenmodes does not rely on random thermal fluctuations to explore their Hilbert space in an inefficient way.   

In conclusion, we have shown that optical eigenmode imaging allows the indirect complex reconstruction of a transmissive target. This method is based on the decomposition of the field into intensity optical eigenmodes. The decomposition coefficients are directly measured by the first-order cross-correlation between the unknown target field  ${\rm T}$ and the optical eigenmodes. Superimposing the optical eigenmodes with the decomposition coefficients yields the reconstructed target field $\rm T$.  Importantly, the complex decomposition coefficients  provide phase-information potentially allowing the reconstruction of optical path length information of the target akin to transmission optical tomography.  Further, we show that the optical eigenmode imaging corresponds to a compressive full-field sampling improving the imaging speed without loss of details enabling fast spectroscopic imaging applications such as Raman imaging. Finally, we used the concept of optical eigenmodes to describe the first-order and second-order correlation functions. In future work we aim to explore the implications of optical eigenmodes in the field of quantum optics.

\begin{acknowledgments}
We thank the UK Engineering and Physical Sciences Research Council for funding. 
ACDL is an EPSRC post doctoral fellow and KD is a Royal Society-Wolfson Merit Award Holder.
\end{acknowledgments}

\bibliography{bib}

\end{document}